# Unlocking Dynamic Luminescent Mapping of pH with Sustainable Lignin-Derived Carbon Dots with Multimodal Readout Capacity

Maja Szymczak[a,*], Jan Hočevar[b], Jernej Iskra[b], Darja Lisjak[c], Jelena Papan Djaniš[b,d,*], Lukasz Marciniak[a,*], Karolina Elzbieciak-Piecka[a,*],

[a] Institute of Low Temperature and Structure Research Polish Academy of Sciences, 50-422 Wroclaw, Poland

[b] Faculty of Chemistry and Chemical Technology, University of Ljubljana, 1000 Ljubljana, Slovenia

[c] Department for Materials Synthesis, Jožef Stefan Institute, 1000 Ljubljana, Slovenia

[d] Centre of Excellence for Photoconversion, Vinča Institute of Nuclear Sciences, National Institute of the Republic of Serbia, University of Belgrade, 11351 Belgrade, Serbia

*corresponding authors: m.szymczak@intibs.pl, jelena.papandjanis@fkkt.uni-lj.si, l.marciniak@intibs.pl, k.elzbieciak@intibs.pl

**Abstract**

In this work, we demonstrate the use of CQDs synthesized from lignin - currently one of the most abundant and underutilized by-products of paper and pulp production - for advanced pH monitoring applications. The presented approach integrates green chemistry principles with an operator-friendly, low-cost, and practical solution for spatial and temporal pH measurement. CQDs functionalized with *m*-aminophenylboronic acid enable highly sensitive and reversible pH readouts through two complementary mechanisms - ratiometric monitoring of emission band intensities, and direct visual observation of colorimetric changes reflected in the CIE1931 chromaticity coordinates. The system achieves maximal sensitivities of 137% $pH^{-1}$ and 49.5% $pH^{-1}$, respectively, while simultaneously maintaining high measurement resolution and full reproducibility of the readouts, placing it among the most effective CQD-based pH sensors reported to date. Here, we demonstrate the capability of 2D luminescent imaging of pH distributions, allowing for both spatially resolved and time-resolved monitoring. Employing just an excitation source, a digital camera or smartphone, and RGB channel analysis, the setup eliminates the necessity for specialized filters or sophisticated instrumentation. The combination of multimodal readout strategies with the capacity for large-area visualization establishes lignin-derived CQDs as a sustainable and practical platform for pH sensing. By simultaneously addressing the challenges of waste valorization and the demand for innovative sensing technologies, this solution fulfills the requirements of both environmentally responsible material design and next-generation pH sensor development.

## 1. Introduction

Discarded cigarette butts, expired pharmaceuticals, and fruit peels are just a few examples of wastes that can be transformed into carbon quantum dots (CQDs), illustrating the potential of converting abundant by-products into functional materials.[1–4] These examples highlight the urgent need to valorize waste streams into functional, value-added materials, thereby contributing to sustainable resource management and reducing reliance on consumable raw materials. In light of the increasing global demand for non-renewable raw materials, the pursuit of sustainable production processes has shifted from being an option to becoming an urgent necessity. A rational and efficient approach to the use of limited resources - whether in laboratory synthesis, industrial manufacturing, or large-scale technological applications - has therefore gained critical importance. The growing awareness of consumerism and the environmental consequences of excessive waste generation has further accelerated initiatives aimed at reducing material consumption and promoting resource circularity, even at the level of routine laboratory practices. As a result, current research increasingly emphasizes strategies for "upcycling" waste streams, including biomass, while simultaneously optimizing processes to ensure more responsible and sustainable material use.[5–8] Within this context, green nanotechnology has emerged as a particularly promising pathway, offering the synthesis of advanced functional nanomaterials from renewable sources as a means to align technological innovation with environmental stewardship.[9–12]

In the field of luminescent materials, the dominant approaches rely on rare-earth and transition-metal dopants. While highly effective in many applications, these materials are inherently limited by the scarcity of critical raw elements, their high extraction costs, and the environmental burden associated with their mining and processing.[13,14] This has driven the search for more sustainable alternatives that can deliver comparable functionality without relying on non-renewable resources. One of the most dynamic and rapidly expanding directions is the development of carbon quantum dots (CQDs) derived from biomass.[15–17] CQDs can be synthesized from a wide variety of renewable precursors, ranging from lignocellulosic biomass and agricultural residues to food industry byproducts.[16,18–21] This versatility in precursor selection not only enables the valorization of abundant waste materials but also allows for scalable, low-cost, and environmentally friendly synthesis routes. Importantly, CQDs exhibit a range of properties that make them exceptionally attractive for luminescence-based applications: their emission can be finely tuned through precursor choice and synthesis conditions; they demonstrate excellent water dispersibility, high biocompatibility, and low cytotoxicity, and offer remarkable chemical and photostability compared to conventional organic phosphors. A further advantage lies in the ease with which their surface chemistry can be tailored, enabling the design of application-oriented nanomaterials with customized functionalities. This adaptability has led to their use across diverse domains, including chemical and biological sensing, bioimaging, light-emitting devices, and photocatalysis. Of particular significance, however, is their growing role in the optical sensing of physical variables, where the broad tunability of their luminescence opens new opportunities for the development of sustainable, highly sensitive, and application-specific sensing platforms.[22–25]

So far, the potential of CQDs for monitoring temperature, humidity, pH, and other physicochemical parameters has been well established.[22–28] Among these, luminescence-based pH monitoring, employing quantum dots or other luminescent materials, offers particular advantages - especially in industrial contexts - over conventional approaches such as electrode-based devices or disposable indicators like pH papers. The key benefit lies in the ability of luminescent probes to provide spatially resolved information, enabling monitoring of pH variations across the entire surface of a monitored object, something unattainable with traditional methods. Likewise, classical molecular indicators, such as phenolphthalein, merely signal the crossing of a certain pH threshold and cannot deliver quantitative, continuous readouts. By contrast, luminescent markers give the possibility of both immediate visual detection through pronounced color changes and precise determination of pH values via calibration of luminescence parameters. This approach not only facilitates remote readout but also enables real-time tracking of pH dynamics, underscoring the strong potential of luminescence-based strategies for practical applications.

To address the possibilities outlined above, this study explores the use of carbon quantum dots synthesized from lignin - a sustainable biomass waste generated during paper production - for pH sensing. The surface chemistry of the CQDs was modified through lignin acidolysis and subsequent *m*-aminophenylboronic acid functionalization.[29] As a result, the modified nanomaterials exhibited a distinct, visually perceptible luminescence color shift - from blue under acidic conditions to green in alkaline environments, which can be attributed to the increasing contribution of surface-state emission at higher pH values. The magnitude of this spectral shift enabled the development of a ratiometric pH sensor based on the intensity ratio of selected spectral ranges of emission bands, providing a usable pH sensing range of approximately 7-10 and exhibiting exceptionally high maximum sensitivity of 137% $pH^{-1}$ at pH~8. In parallel, a second pH readout mode based on changes in CIE1931 chromaticity coordinates was also proposed, achieving a sensitivity of 49.5% pH$^{-1}$. However, both these approaches require the recording of emission spectra of the analyzed CQDs. The central advance of this work is the proof-of-concept demonstration that the designed sensor enables dynamic, real-time two-dimensional (2D) pH mapping across large areas using only a UV excitation source and a conventional digital camera by the analysis of the luminescence intensity ratio recorded in its RGB channels. This study provides the evidence that luminescent pH mapping can be performed without additional optical filters, overcoming a key limitation of conventional ratiometric luminescent imaging. In traditional setups, the necessity of employing optical filters not only increases the complexity and cost of the instrumentation but also introduces practical drawbacks. Filter exchange is inherently time-consuming and may interrupt continuous measurements, creating the risk of changes in the monitored parameter during the switching process. Such interruptions are particularly problematic for highly dynamic systems, where rapid pH fluctuations can occur on timescales shorter than the measurement delay. By eliminating the need for filters, the proposed approach ensures uninterrupted monitoring, significantly simplifies the optical setup, and enhances both the temporal resolution and accessibility of luminescent pH imaging. Accordingly, this study demonstrates that real-time, dynamic pH monitoring can be achieved through a simple imaging strategy based on RGB channel extraction from digital photographs, followed by their intensity ratio analysis. This straightforward and filter-free approach, implemented using only a consumer-grade camera, transforms widely accessible imaging tools into effective luminescent sensing platforms. Moreover, the use of sustainable, lignin-derived CQDs for pH imaging establishes a new direction in the development of luminescent sensors that combine environmental responsibility with simplicity, affordability, and scalability. This strategy holds great promise for future applications in environmental monitoring, biomedical diagnostics, and lab-on-chip technologies, where accessible, low-cost, and real-time pH mapping is of critical importance.

## 2. Experimental Section

*Materials*

The used Kraft lignin was manufactured by Lignocity from Sweden; *m*-aminophenylboronic acid was purchased from Fluorochem, hydrochloric acid (37% aq. sol.) was purchased from Pregl Chemicals, while sodium hydroxide (p.a.) was purchased from SigmaAldrich.

*Synthesis*

Carbon quantum dots derived from lignin were synthesized via a two-step hydrothermal procedure.[29] In the first step, 0.3 g of spruce-derived lignin was dispersed in water together with *m*-aminophenylboronic acid and stirred at 90 °C for 1 h. After cooling to room temperature, the suspension was filtered through a 0.45 µm membrane and transferred into a 50 cm$^3$ Teflon-lined autoclave, followed by hydrothermal treatment at 200 °C for 12 h. The resulting solution was subsequently filtered through a 0.22 µm membrane, and the obtained filtrate was purified by dialysis (molecular weight cut-off: 500-1000 Da) in deionized water for 48 h. The resulting colloid (denoted NBCQD) was stored at 4 °C until further use.

For the preparation of acid-pretreated samples, the initial step involved the addition of 20 µL (NBCQD-20) or 1000 µL (NBCQD-1000) of HCl to the aqueous lignin suspension, followed by mixing at 60 °C for 30 min. The subsequent steps of the synthesis were performed analogously to the procedure described for NBCQD.

*Characterization*

Infrared spectra were recorded between 4000 cm$^{-1}$ and 400 cm$^{-1}$ using a Bruker Alpha-II FT-IR spectrophotometer with ATR module. Raman spectra were recorded with Bravo Handheld Raman Spectrometer, Bruker Optic. Electro-kinetic measurements (Zeta potential) of QDs dispersed in double deionized water with a concentration of 1 mg ml$^{-1}$ and at 25 °C were performed using a Litesizer 500 from Anton Paar. The pH was adjusted with HCl and NaOH solutions (0.1 or 1 M). 2D-HSQC NMR spectra were recorded at 25 °C using a Bruker Avance III 500 spectrometer. The following acquisition parameters were used: F2 range = 10 to 0 ppm, F1 range = 158 to –8 ppm, number of scans (ns) = 24, dummy scans (ds) = 16, number of increments (ni) = 256, relaxation delay (d1) = 1.47 s, and pulse program = hsqcetgpsi2. The spectra were processed with Bruker TopSpin 4.1.1 and MestReNova software. Morphology of QDs was controlled by the use of transmission electron microscope (TEM) Jeol 2100, Tokyo, Japan. Colloids were drop-deposited on the Cu-supported TEM grid and left to dry. The nanoparticle size was determined from TEM images using ImageJ software. Excitation and emission spectra as well as luminescence decay profiles as a function of pH were recorded using FLS 1000 Fluorescence Spectrometer from Edinburgh Instruments equipped with a R928P side window multiplier tube from Hamamatsu as a detector as well as a 450 W xenon lamp and picosecond pulsed light source AGILE (also from Edinburgh Instruments) as the excitation sources. The temperature during the measurements was externally controlled using a Quantum Northwest TC1 Temperature Controller with associated Peltier-controlled cuvette holder. The pH value of colloids was controlled using a Seven Compact S210 pH meter form Mettler Toledo. To change the pH, solutions of HCl and NaOH were used. Luminescence properties were measured in the pH range from approximately 2 to 12 by gradually adding a defined volumes of NaOH/HCl solution.

The average lifetime of the excited state ($\tau_{avr}$) was determined based on biexponential fit of decay curves using equations:

$$\tau_{avr} = \frac{A_1\tau_1^2 + A_2\tau_2^2}{A_1\tau_1 + A_2\tau_2} \qquad (1)$$

$$I(t) = I_0 + A_1 \cdot \exp\left(-\frac{t}{\tau_1}\right) + A_2 \cdot \exp\left(-\frac{t}{\tau_2}\right) \qquad (2)$$

where $\tau_1$ and $\tau_2$ are the decay components and $A_1$ and $A_2$ are the amplitudes of the biexponential function.

Photographs for the proof-of-concept experiment, as well as for colloidal samples upon daylight and Xenon lamp excitation ($\lambda_{exc}$ = 400 nm, provided by the Xenon lamp coupled to the FLS spectrometer), were taken using a Canon EOS 400D camera equipped with an EFS 60 mm macro lens and an appropriate long-pass optical filter to eliminate the excitation beam. In the proof-of-concept experiment, a crystal of NaOH was added to a droplet of the acidic colloid of CQDs. During the NaOH crystal dissolution process, photographs were acquired over time with a camera integration time of 1.3 s. The blue and green (RGB) channels were extracted from the photographs using IrfanView 64 4.51 software. Subsequently, the B and G intensity maps were processed and divided by each other using ImageJ 1.8.0_172 software.

## 3. Results and discussion

Understanding the morphology of the synthesized carbon dots is essential, as their size and shape strongly influence the resulting optical properties. Therefore, the obtained samples, namely NBCQD, NBCQD-20, and NBCQD-1000, were examined with respect to their morphological characteristics (Figures 1a-c). TEM analysis revealed nanometric particles with a predominantly spherical shape. Although some agglomeration was observed, this effect can be attributed to the drying of colloidal dispersions required for TEM sample preparation. Particle size distributions obtained from TEM images revealed a clear effect of lignin pretreatment on the morphology of the resulting carbon dots. In the absence of HCl, the average particle size was 25.8 nm, whereas pretreatment with acid led to significantly smaller nanoparticles, with average sizes of 3.9 nm for NBCQD-20 and 1.4 nm for NBCQD-1000 (Figures 1d-f). The reduction in particle size with increasing HCl content can be ascribed to the more pronounced degradation and fragmentation of the lignin framework under strongly acidic conditions, which favors the nucleation of smaller carbon domains.[29]

Furthermore, HSQC NMR measurements were performed with particular emphasis on the aromatic region (Figure 1h-j). NMR analysis of lignin quantum dots obtained from kraft spruce lignin without the use of HCl in the pretreatment phase revealed that the aromatic structure remains relatively unaltered, and with weaker signals from the aromatic nuclei relative to acid-treated samples (Figure 1g). The use of 20 µL HCl in the pretreatment step led to LCQD with four distinct aromatic nuclei at (7.1; 121.1), (7.3; 121.1), (7.4; 125.1), and (7.5; 129.3) in the 2D-HSQC NMR spectra (Figure 1h). These resonances show downfield shifts in the proton dimension compared to the untreated sample, while the corresponding carbon chemical shifts remain largely unchanged. This pattern suggests that the aromatic ring undergoes only partial modification, with changes localized to the electronic environment surrounding the protons rather than the carbons. At a higher HCl concentration (1000 µL), the carbon dimension of aromatic chemical shifts also shows a substantial downfield shift, with nuclei detected at (7.4; 125.2), (7.4; 129.5), (7.7; 128.9), and (7.8; 133.7) (Figure 1i). These pronounced carbon shifts indicate the incorporation of electron-withdrawing substituents, which pull electron density away and deshield carbon or proton atoms, resulting in larger shifts in the HSQC NMR spectrum. Since XPS analysis has shown the presence of Cl on the surface of CQDs, the latter could be a counterion of the protonated amino group, which is also a strong electron-withdrawing group. Taken together, these observations indicate that increasing the HCl concentration promotes the incorporation of -$NH^{3+}$ and $Cl^-$ groups, leading to progressively more pronounced downfield shifts in both the 1H and 13C dimensions. The XPS analysis further supports this conclusion, showing that higher HCl concentrations correlate with an increased proportion of nitrogen in the form of quaternary ammonium (-$NR^{3+}$) species.[29]

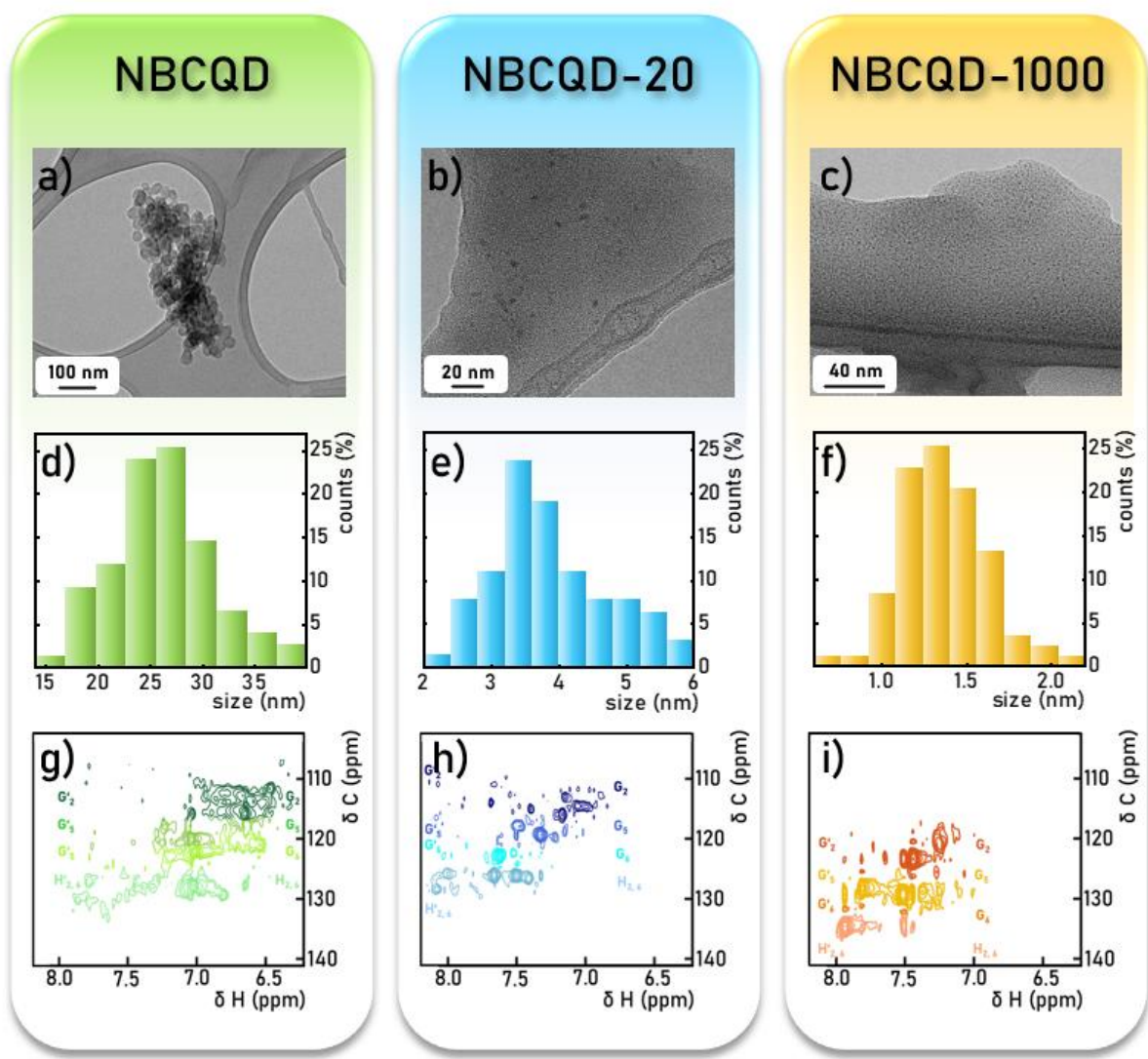

**Figure 1.** Representative TEM images (a–c), corresponding particle size distribution histograms (d–f), and 2D-HSQC NMR spectra (g–i) for NBCQD (a, d, g), NBCQD-20 (b, e, h), and NBCQD-1000 (c, f, i) samples.

To further assess the structural characteristics, Raman spectroscopy was employed, as it provides insight into the degree of disorder and the extent of graphitization in carbon-based materials. Particular attention was paid to the relative intensities of the two characteristic Raman bands: the D band, attributed to disordered $sp^3$-hybridized carbon at defect sites and grain boundaries, and the G band, corresponding to the stretching vibrations of $sp^2$-hybridized carbon atoms within graphitic domains.[30,31] For all NBCQDs, the Raman spectra exhibited well-defined D and G bands at approximately 1430 and 1640 cm$^{-1}$, respectively (Figure 2a). A systematic decrease in the $I_D/I_G$ ratio was observed, with values of 0.78, 0.61, and 0.42 for NBCQD, NBCQD-20, and NBCQD-1000, respectively. This decreasing trend indicates a gradual increase in structural ordering and graphitization as the amount of HCl used in lignin pretreatment increases, highlighting the pivotal role of acid concentration in governing the structural evolution of the resulting CQDs. A higher acidity level likely facilitates more efficient hydrolysis and depolymerization of lignin, leading to the removal of amorphous components and promoting the formation of graphitic domains with improved ordering. Consequently, the CQDs obtained under strongly acidic conditions exhibit a more crystalline structure.

The obtained NBCQD colloids were also examined by FTIR spectroscopy to identify the surface functional groups and to confirm the successful modification of the CQDs with *m*-aminophenylboronic acid. The FTIR spectra revealed a set of absorption bands extending from approximately 1000 to 1600 cm$^{-1}$, characteristic of vibrations associated with aromatic moieties typically present in carbon dots (Figure 2b). In particular, the band at 1603 cm$^{-1}$ can be assigned to C-O stretching vibrations, those around 1050 cm$^{-1}$ correspond to in-plane C-H bending and band near ~1350 cm$^{-1}$ is attributed to C=C stretching vibrations within aromatic rings. In addition, a distinct band at ~1710 cm$^{-1}$ was observed, which is characteristic of C=O stretching vibrations, indicating the presence of oxygen-containing surface functionalities. Moreover, the recorded spectra clearly revealed the presence of boron-containing groups, as evidenced by characteristic absorption bands at 1029 cm$^{-1}$, 1086 cm$^{-1}$, and 1342 cm$^{-1}$, assigned to B-O-H, B-O, and B-N stretching vibrations, respectively. These bands confirm the covalent attachment of boronic functionalities onto the CQD surface. A broad band centered at around 3210 cm$^{-1}$, corresponding to N-H stretching vibrations, further supporting the presence of surface amino groups. On the other hand, a band at 699 cm$^{-1}$, ascribed to C-Cl stretching vibrations, appears in the NBCQD-20 and NBCQD-1000 samples. Its appearance indicates the incorporation of chlorine-containing functional groups onto the QDs surface.[32,33] The presence of these features provides evidence of the effective surface functionalization, which is expected to play a key role in tailoring the interaction of the CQDs with protons and thus in enabling their pH-sensing functionality. These findings are further supported by Zeta potential measurements, which are discussed in the following section.

Zeta potential measurements were conducted on the obtained colloids to complement the above-described analyses and to gain insights into the surface chemistry of the quantum dots, particularly the role of surface ligands and their behavior with pH. Although the absolute values of ζ-potential do not provide a direct identification of specific ligands, when measured under identical conditions of temperature and concentration and other relevant factors, which were carefully controlled in the present study, they can yield relative information on the abundance of charged surface groups.[34,35] Specifically, more negative values suggest a higher density of negatively charged ligands, while less negative (or more positive) values indicate either an increased contribution of positively charged groups or a reduction in negatively charged functionalities. As shown in Figure 2c, samples synthesized with lignin acidification consistently exhibit higher zeta potential values across the investigated pH range, with the effect becoming more pronounced as the amount of acid used increases. This trend implies the introduction of additional positively charged surface ligands (e.g., protonated amino groups, -$NH_3^+$), a decrease in the number of negatively charged groups ($OH^-$), or a combined effect of both processes.[36,37] Taken together with the results of FTIR and NMR analyses, these findings enabled to propose a model of the ligand arrangement on the surface of the NBDQDs, schematically illustrated in Figure 2e. In general, based on the experimental results and literature reports, several conclusions can be drawn regarding the impact of lignin pretreatment with HCl: (1) enhanced incorporation of nitrogen originating from *m*-aminophenylboronic acid into the graphitic structure; (2) more effective deposition of $B(OH)_2$ groups on the surface at the expense of hydroxyl groups; (3) attachment of chloride groups, whose concentration increases with the amount of HCl used; and (4) a higher density of amino functionalities on the surface. In general, all these effects can be attributed to the fact that acidic pretreatment with HCl partially removes or protonates native hydroxyl groups in lignin, thereby generating more reactive sites for subsequent attachment of various surface moieties.[29]

Beyond providing information on surface charge, zeta potential measurements also serve as an indicator of colloidal stability. As a general criterion, dispersions are considered electrostatically stable when their ζ-potential values fall below -30 mV or exceed +30 mV.[34,35] In the present case, the stability of NBCQDs was found to be strongly governed by the degree of acid pretreatment od lignin applied during synthesis. The NBCQD colloid exhibited values below -30 mV over a broad pH range, extending from pH~6 upwards, indicating favorable stability (Figure 2c). A narrower stability window, from approximately pH~8 and above was observed for NBCQD-20, while NBCQD-1000 does not reach values lower than -30 mV within the investigated pH range. Hence, we can conclude that with HCl pretreatment, the concentration of available, protonatable amine groups on the surface is increased. This is in accordance with the XPS spectra[29], and it simultaneously highlights the tunability of surface charge through controlled lignin pretreatment. It is also worth mentioning that the obtained Zeta potential values are comparable to, or in some cases even exceed, those reported for other functionalized CQDs reported so far.[38–40] This suggests that, despite the observed variations, the overall colloidal stability of the investigated systems falls within the range commonly considered acceptable for CQDs and is sufficient for practical use in short-term sensing and optical experiments. Importantly, visual observations and captured photographs (Figure 2d and S1) demonstrated that the studied colloids remained well-dispersed across the entire investigated pH range, without any signs of precipitation, aggregation, or sedimentation. This persistence of colloidal stability under both acidic and alkaline conditions underscores the robustness of the synthesized quantum dots and highlights their practical suitability for sensing applications, where reliable performance is required in dynamically changing chemical environments.

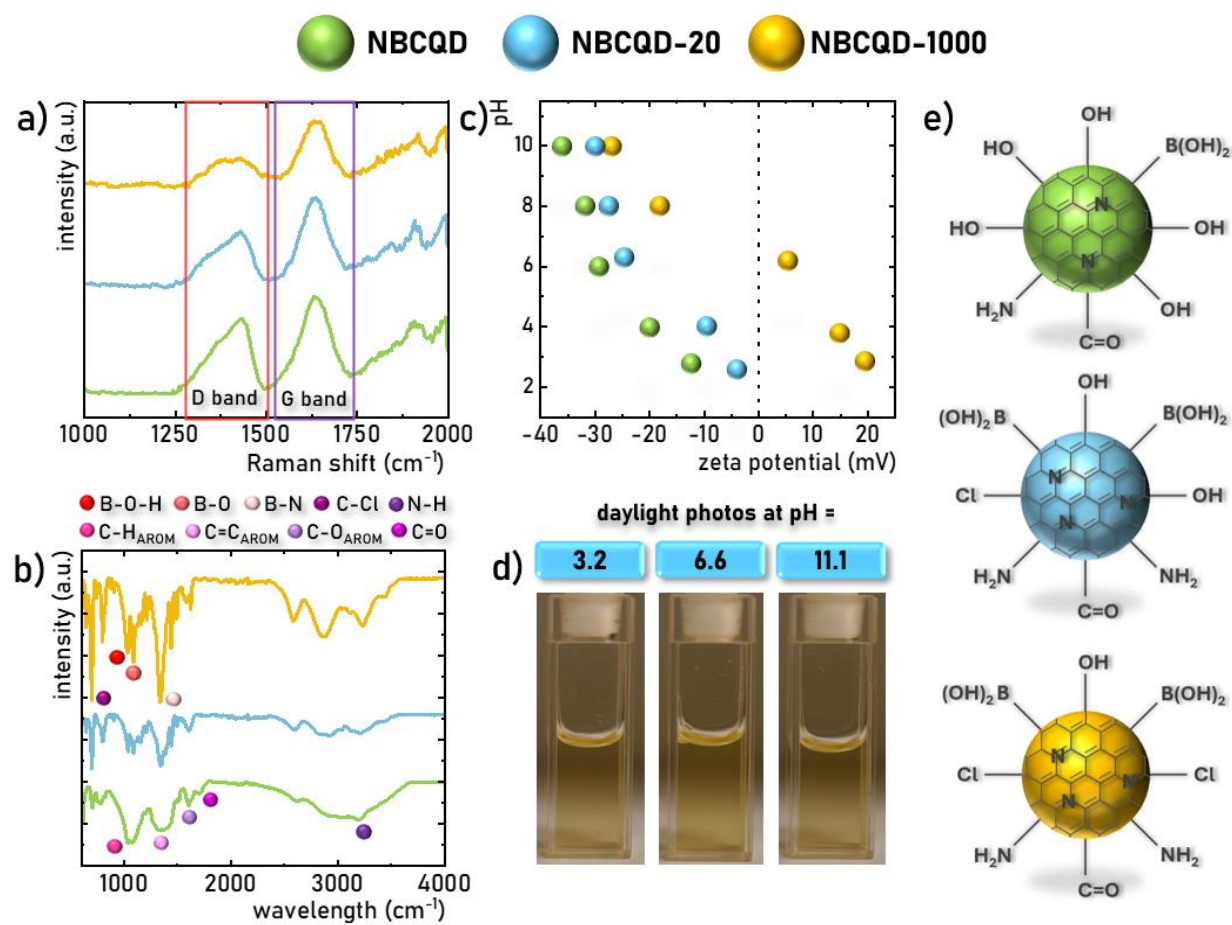

**Figure 2.** Room-temperature Raman spectra (a), FTIR spectra (b), and Zeta potential as a function of pH (c) for NBCQD, NBCQD-20, and NBCQD-1000. Daylight photographs of NBCQD-20 colloid at pH = 3.2, 6.6, and 11.1, (d). Schematic visualization of NBCQDs' surface (e).

The luminescence of carbon dots is known to be highly sensitive to a variety of factors, both internal and external. Previous reports have demonstrated that their optical behavior, including emission color, can be effectively tuned through modulation of particle size and morphology, alteration of surface ligands, or incorporation of heteroatoms into the graphitic framework.[41–45] Such versatility makes carbon dots an attractive platform for luminescence-based technologies; however, it also highlights the need for a comprehensive understanding of the interplay between their structure and photophysical response. A thorough investigation of these relationships is essential to enable the rational design of reliable, application-oriented systems with predictable and controllable emission characteristics. In this context, the NBCQDs colloids synthesized in this work were subjected to detailed spectroscopic characterization to elucidate their luminescent properties and establish correlations with their structural features. To ensure reliable comparison, all measurements were performed under constant conditions (pH = 7, T = 20 °C). Figure 3b shows a luminescence map of normalized emission spectra for NBCQD-20 recorded upon different excitation with wavelengths from 260 to 490 nm. At shorter excitation wavelengths (260-380 nm), the spectra are dominated by a broad emission band centered at 422.5 nm, which can be attributed to the graphitic carbon dot core and is associated with π*→π electronic transitions in $sp^2$-hybridized carbon domains. When the excitation wavelength is increased to 400 nm, a new broad emission band emerges at approximately 500 nm, marking a transition in the emission behavior. Further red-shifting the excitation to 420 nm results in the complete quenching of the π*→π emission band, leaving only the emission band centered at 520 nm. This green emission is most likely linked to inter-state electron transitions (n*→π) mediated by the coordination of surface ligands, in particular N- and B-containing groups or carbonyl (C=O) functionalities, whose presence was confirmed by earlier FTIR, NMR and XPS analysis. Such surface moieties generate multiple emissive trap states between the HOMO and LUMO molecular orbitals, characterized by narrower energy gaps and, consequently, by longer-wavelength emission. The simplified scheme in Figure 3a illustrates only two dominant types of electronic transitions; however, the distinctly non-Gaussian shape of the emission bands indicates the coexistence of numerous trap states, each contributing to the complex optical behavior of the material. [45–47]

Another effect observed in the emission spectra as a function of excitation wavelength manifests as a systematic shift of the emission maxima with varying excitation energy. Two opposite trends are evident: the π*→π band undergoes a red shift with increasing excitation wavelength, whereas the n→π band exhibits a blue shift. Excitation-dependent emission shifts are a well-documented phenomenon in quantum-dots-related literature[48], although the underlying mechanisms are often difficult to unambiguously confirm. In general, two primary explanations are considered: (1) heterogeneity in nanoparticle size within the colloid and (2) the presence of multiple surface-related trap states. In the first case, the shift arises from size-dependent band-gap energies, where larger quantum dots possess smaller band gaps and therefore emit at longer wavelengths, leading to a red shift as the excitation wavelength increases. In the second case, sequential activation of emissive trap states with progressively lower energies accounts for the spectral red shift. Both mechanisms are consistent with the behavior of the shorter-wavelength emission band in NBCQD colloids, which likely reflects a synergistic contribution of size heterogeneity (as evidenced by TEM) and the activation of surface trap states related to the presence of ligands. Additional factors may also contribute to the observed trends. These include excitation-energy-dependent energy transfer between emissive trap states, redistribution of charge carriers among surface defect sites, or competition between radiative and non-radiative pathways that become selectively favored at different excitation energies.[49–51] The combination of these processes underscores the complexity of the photophysics of carbon quantum dots, where emission rarely originates from a single well-defined transition, but rather from a dynamic interplay of core states, surface traps, and ligand-related fluorophores.

Excitation spectra were measured by monitoring emissions at 425 and 515 nm, revealing excitation bands distributed over a broad spectral range (Figure 3c). In both cases, a dominant band with a maximum at approximately 370 nm was observed, corresponding to π→π* transitions in $sp^2$-hybridized carbon domains. For the spectrum monitored at $\lambda_{em}$ = 515 nm, this band is noticeably broadened, indicating an additional component and suggesting the involvement of multiple emission centers.[52,53] Such broadening probably arise from defect-related states introduced by heteroatom doping - in this case, nitrogen. Distinct higher-energy bands are also observed at ~260 nm and ~300 nm, both directly linked to surface chemistry. The ~300 nm band can be primarily assigned to n→π* transitions of C=O groups, while contributions from B(OH)$_2$ functionalities are also expected. Same for a band at ~490 nm ($\lambda_{em}$ = 515 nm). They may be associated with charge-transfer (CT) processes, either between surface ligands (e.g., NH$_2$ acting as an electron donor and C=O as an acceptor) or between surface ligands and the carbon core. Alternatively, this feature may originate from deep surface states generated by strongly interacting or specifically arranged functional groups, particularly C=O, B(OH)$_2$, OH and NH$_2$ groups acting as donors and acceptors.

Time-resolved photoluminescence of the analyzed CQDs was measured at $\lambda_{em}$ = 425 and 515 nm (Figure 3d) to analyze the dynamics of the two emissions - the blue "carbon-core-state" band and the green "surface-state" band. Both profiles were fitted with a biexponential model according to Eq. 2. For $\lambda_{em}$ = 425 nm, the decay was effectively single-exponential, with the second component contributing negligibly, resulting in an average excited-state lifetime of 3.14 ns (calculated accordingly to Eq. 1). This behavior indicates that the 425 nm emission originates predominantly from uniform and well-passivated *sp$^2$*-carbon-core states, characterized by limited access to non-radiative pathways. On the other hand, the decay curve obtained at $\lambda_{em}$ = 515 nm exhibited a pronounced multi-exponential character, reflecting the heterogeneous nature of emissive surface states. The fit revealed a dominant fast component, accompanied by a weaker long-lived contribution, giving an average lifetime of 0.75 ns. The reduced lifetime associated with the surface-state emission can be ascribed to enhanced non-radiative deactivation channels. These include stronger coupling of surface-localized states to high-energy vibrational modes such as -OH and -NH$_2$ groups, which promotes multiphonon relaxation, as well as efficient charge or energy transfer to nearby defect sites. Consequently, the emission at 515 nm is more strongly quenched compared to the core-related band at 425 nm.

The luminescence properties of NBCQD, NBCQD-20, and NBCQD-1000 colloids were systematically compared under identical conditions - pH = 7 and T = 20 °C (Figure 3e). For all samples, the emission spectra showed a dominant broad π→π* band from $sp^2$-hybridized carbon domains at ~422.5 nm. While the spectra of NBCQD and NBCQD-1000 were largely comparable, NBCQD-20 exhibited a relatively stronger contribution of the 515 nm band, indicative of enhanced emission from defect-related surface states. In contrast, the excitation spectra revealed notable differences: NBCQD-20 and NBCQD-1000 displayed a greater contribution from high-energy excitation bands compared to NBCQD (Figure 3f). This feature is consistent with the interaction of $B(OH)_2$ ligands, which are more abundant in samples prepared with HCl lignin pretreatment (NBCQD-20 and NBCQD-1000).

When comparing the luminescence kinetics of the carbon-core emission at 425 nm across all samples (Figure 3g), the longest average lifetime was observed for NBCQD-20, reaching 3.14 ns. A shorter value of 2.48 ns was obtained for NBCQD-1000, while the fastest decay was observed for NBCQD with a lifetime of 2.12 ns (Figure 3h). These variations can be explained by considering both the relative contribution of surface-related emission and the chemical nature of the functional groups on the nanoparticle surface. In the case of NBCQD-20, the prolonged lifetime is consistent with the higher contribution of the green band at 515 nm, which is associated with defect-related surface states. The coexistence of both core and defect emissive channels may slow down the apparent decay dynamics by redistributing excitations between different sites. Moreover, the surface composition of NBCQD-20, characterized by a higher content of -Cl and -$NH_2$ groups at the expense of hydroxyl functionalities, compared to NBCQD, likely provides more effective surface passivation and reduces non-radiative deactivation through -OH vibrations. By contrast, the mentioned NBCQD contains the highest density of hydroxyl groups, which are known to couple strongly to high-energy vibrational modes, thereby facilitating multiphonon relaxation and resulting in a shorter excited-state lifetime. In turn, NBCQD-1000 exhibits an intermediate behavior, with a reduced number of hydroxyl groups compared to NBCQD but a surface composition comparable to NBCQD-20. This partial decrease in quenching -OH groups explains the moderately longer lifetime relative to NBCQD, though the effect is less pronounced than in NBCQD-20. Altogether, these results highlight the critical role of surface functionalization in controlling the balance between radiative and non-radiative pathways in carbon quantum dots, directly governing their excited-state dynamics.

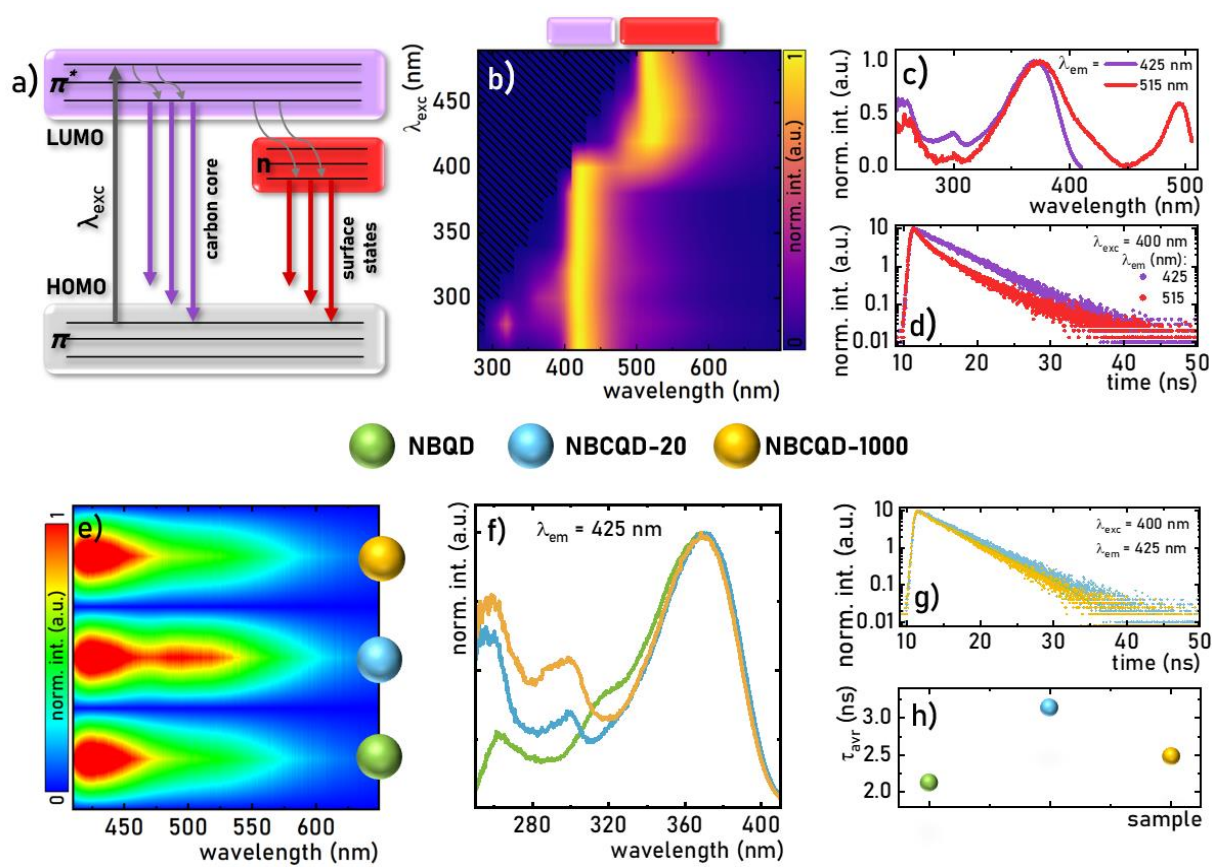

**Figure 3.** Schematic illustration of the energy levels and electronic transitions in NBCQD materials (a). Luminescence map of normalized emission spectra as a function of excitation wavelength (b), excitation spectra recorded at $\lambda_{em}$ = 425 and 515 nm (c) and luminescence decay profiles ($\lambda_{exc}$ = 400 nm; $\lambda_{em}$ = 425 and 515 nm) (d) for colloidal NBCQD-20 (pH = 7, T = 20 °C). Maps of normalized emission spectra (e), normalized excitation

spectra (f), luminescence decay curves (g), and corresponding calculated average lifetimes τ$_{avr}$ of excited state (h) for colloidal NBCQD, NBCQD-20, and NBCQD-1000 (pH = 7, T = 20 °C).

It is well established that the luminescence of carbon quantum dots is highly sensitive to variations in physicochemical parameters, with pH being among the most influential.[54–56] To investigate this effect, the luminescence response of NBCQDs colloids to pH changes upon λ$_{exc}$ = 400 nm was examined under strictly controlled conditions at 20 °C. For all three colloidal samples, emission spectra were recorded over the pH range of approximately 2.5–12 (Figure 4a-c and S2-S4) The analysis revealed two distinct spectral behaviors: (1) emission dominated by the π*→π electronic transitions of *sp$^2$*-hybridized carbon domains in the core, with a maximum at ~422 nm, and (2) emission spectrum dominated by the band associated with n*→π electronic transitions associated with surface trap states. The former band was predominant for all colloids up to pH~8. At higher pH values, both bands appeared simultaneously, and beyond pH~8.5 the spectra were dominated by the surface-state emission band, which remained prevalent throughout the alkaline range investigated.

In general, the literature describes numerous mechanisms responsible for the pH-dependent changes in the luminescence of carbon quantum dots, yet none of them has been identified as the sole "culprit" for the observed spectral effects, nor has any of them been conclusively confirmed.[54,57] Instead, these processes are usually regarded as hypothetical and strongly interdependent, with the experimentally observed changes most often resulting from their cumulative effect. Among the most frequently mentioned factors are pH-induced modifications of electronic energy levels, protonation and deprotonation of functional groups, aggregation of quantum dots, and electron or energy transfer between dots and/or ligands. From the perspective of NBCQDs colloids, the surface chemistry of the nanoparticles probably plays a particularly critical role. The surface hosts a rich variety of functional groups and defect states, which not only influence the intrinsic radiative processes within the carbon core but also govern interparticle interactions in aqueous solutions. For instance, a change in surface charge alters the colloidal stability, potentially leading to aggregation, or modifies dipole-dipole interactions between surface groups, which can introduce additional non-radiative channels or energy-transfer pathways. This example highlights the strong interdependence of the aforementioned mechanisms.

Focusing on protonation-deprotonation processes, the most relevant ligands in NBCQDs are -B(OH)$_2$, -NH$_2$, and -OH groups. At elevated pH, deprotonation of the hydroxyl groups, also in the boronic groups, occurs, resulting in negatively charged -O$^-$ species. The formation of such anionic species modifies the electronic environment of the nanoparticle: the negative charge is delocalized over electronegative atoms, altering the electronic density near defect states and thereby shifting the relative positions of the HOMO and LUMO energy levels. Such modifications can reduce or increase the effective bandgap and open additional radiative or non-radiative relaxation channels, manifesting as spectral shifts or intensity changes in luminescence. At low pH values, in particular, -NH$_2$ groups may be protonated to -NH$_3^+$, introducing localized positive charges on the surface. This protonation increases Coulombic interactions and changes the polar environment, which in turn may quench or enhance specific emissive states, depending on whether they are stabilized or destabilized by the altered charge distribution. In a case of Cl$^-$ ligands, Cl$^-$ binding to CQDs depends on surface groups and pH. At low pH, protonated amino groups (-NH$_3^+$) strongly attract Cl$^-$, while -OH and C=O provide weak hydrogen bonding. At neutral pH, binding is mainly via hydrogen bonds, and at high pH, neutral amino groups and negatively charged boronic acid reduce or repel Cl$^-$, resulting in weak overall binding. The -B(OH)$_2$ moiety deserves special attention due to its strong pH responsiveness. At neutral pH, boron is typically *sp$^2$*-hybridized, forming a trigonal planar configuration with hydroxyl groups. Under basic conditions, however, boron can act as a Lewis acid, accepting an additional electron pair from the surrounding hydroxide ions and converting into a tetrahedral *sp³*-hybridized boronate species. This structural transformation substantially alters the electronic properties of the surface, disrupting conjugation with the carbon core and introducing new charge-transfer pathways between boron states and the π-system of the CQDs. Such modifications can significantly influence both the intensity and wavelength of emission.

Taken together, these considerations clearly demonstrate that the observed luminescence variations cannot be ascribed to a single mechanism. Instead, they emerge from a complex interplay of protonation-deprotonation equilibria, charge redistribution, changes in orbital energies, surface defect passivation, and interparticle interactions.

Such pronounced spectral changes enabled the development of ratiometric pH sensors whose response is based on the luminescence intensity ratio between two emission bands, namely:

$$LIR = \frac{\int_{530nm}^{540nm} I(surface\ states) d\lambda}{\int_{410nm}^{420nm} I(carbon\ core) d\lambda} \quad (3)$$

The dependence of emission band intensity within the spectral ranges used to construct the LIR-based calibration curve of pH - namely, 530-540 nm and 410-420 nm - is presented in Figures 4d and 5e, respectively. The emission intensity in the 530-540 nm range decreases with increasing pH from acidic values up to approximately 6.5 for all NBCQD colloids. This decrease is most pronounced for NBCQD-1000, whereas for the other colloids, the change is negligible. Mentioned value represents the isoelectric point for NBCQD-1000 (Figure 2c) where the total electric charge in the system is 0). This luminescent behavior is direct evidence of how the aforementioned mechanisms (protonation-deprotonation, charge redistribution, etc.) can significantly influence the intensity of emission. Beyond this pH point, a sharp increase in intensity is observed up to around pH 8, after which the signal stabilizes and remains constant for all samples. In contrast, the emission intensity in the 410-420 nm range follows the opposite trend: a slight increase is observed up to pH ~8, followed by a rapid decrease at higher pH values. This behavior is particularly evident for NBCQD-20, where the emission intensity drops by nearly threefold between pH 8 and 10. These contrasting spectral behaviors result in a strongly pH-dependent LIR (Figure 4f), with the most pronounced changes occurring between pH~7 and 10. For example, the LIR for NBCQD-20 increases by a factor of 11 between pH 6.6 and 9.3, demonstrating the high dynamic response of the system in this range. A key parameter describing the performance of luminescent sensors is the relative sensitivity ($S_R$), which is calculated according to the following equation:

$$S_R = \frac{1}{LIR}\frac{\Delta LIR}{\Delta pH}100\% \qquad (4)$$

where $\Delta LIR$ denotes the change in the luminescence intensity ratio per unit change in $\Delta pH$. The developed pH sensor exhibits a distinct threshold in sensitivity (Figure 4g), with $S_R$ values increasing markedly above pH~6, indicating enhanced detection capabilities in the neutral to alkaline region. Specifically, the relative sensitivity remained below 1% $pH^{-1}$ at lower pH values and consistently exceeded this threshold above pH 6, marking the onset of the sensor's optimal operating range. The maximum sensitivities reached 137% $pH^{-1}$ at pH = 8.16 for NBCQD, 122% $pH^{-1}$ at pH = 8.00 for NBCQD-20, and 120% $pH^{-1}$ at pH = 8.40 for NBCQD-1000. Around pH 8, all three developed pH sensors exhibit the highest sensitivity due to an optimal balance between surface charge and electrostatic interactions. Under these conditions, the sensor surfaces become highly negatively charged as both hydroxyl and boronic acid groups are deprotonated, while the amino groups remain in their neutral form. This combination maximizes charge repulsion and enhances the sensor's response efficiency.

This exceptionally high sensitivity and the pH range of 6-11 renders the developed system particularly suitable for applications requiring precise detection of small pH shifts within this window. These include physiological processes occurring near neutral pH, quality control in the food and beverage industry, environmental water monitoring, and pH regulation in industrial bioreactors.

Sensitivity is not the only parameter that determines the overall performance and practical applicability of the designed luminescent pH sensor. While high sensitivity ensures a pronounced luminescence response to small changes in pH, it must be complemented by sufficient pH resolution - that is, the smallest detectable change in pH that can be reliably distinguished by the system. High resolution is essential in applications where subtle pH fluctuations carry significant implications, such as in physiological monitoring, enzymatic activity tracking, or fine-tuned industrial processes. Therefore, a comprehensive evaluation of the sensor's performance must consider not only the magnitude of the signal change (sensitivity), but also its precision in discriminating between closely spaced pH values. Therefore, the uncertainty of the developed pH meters based on NBCQD, NBCQD-20, and NBCQD-1000 colloids was determined using the following formula:

$$\delta pH = \frac{1}{S_R} \frac{\delta LIR}{LIR} \qquad (5)$$

where $\delta LIR/LIR$ represents the relative uncertainty in pH determination and is governed by the uncertainties in the quantification of the individual emission intensities ($I_1$ and $I_2$) used to calculate the LIR, as expressed by the following equation:

$$\frac{\delta LIR}{LIR} = \sqrt{\left(\frac{\delta I_1}{I_1}\right)^2 + \left(\frac{\delta I_2}{I_2}\right)^2} \qquad (6)$$

The calculated uncertainty values remained consistently below 0.01 in the pH range of ~7-10, coinciding with the region of maximum sensitivity and thereby defining the effective operating window of the sensors. This level of performance underscores their ability to accurately discriminate even minute pH fluctuations within the optimal range (Figure 4h). Moreover, such low uncertainty highlights the potential of the developed system for applications requiring precise detection of subtle or spatially localized pH variations.

The reversibility of pH-induced luminescence changes was evaluated for the designed pH meter. Reversibility is a critical characteristic for practical sensing applications, as it ensures that the sensor is not limited to single use after exposure to a wide pH range. Instead, it can function as a durable, long-life sensor, significantly reducing operating costs and improving applicability in real-world monitoring. To assess this property, emission spectra were recorded over 7 consecutive cycles of increasing and decreasing the pH between 3.2 and 10.8. The consistent LIR values obtained across these cycles confirmed the excellent reproducibility and stability of the sensor's response (Figure 4i).

The combination of high sensitivity, fine pH resolution, and excellent reproducibility demonstrates that the NBCQD-based pH meter not only ensures reliable and accurate readings, but also exhibits long-term stability and resistance to degradation, making it a highly promising candidate for next-generation pH sensing in environmental monitoring, biomedical diagnostics, and industrial process control.

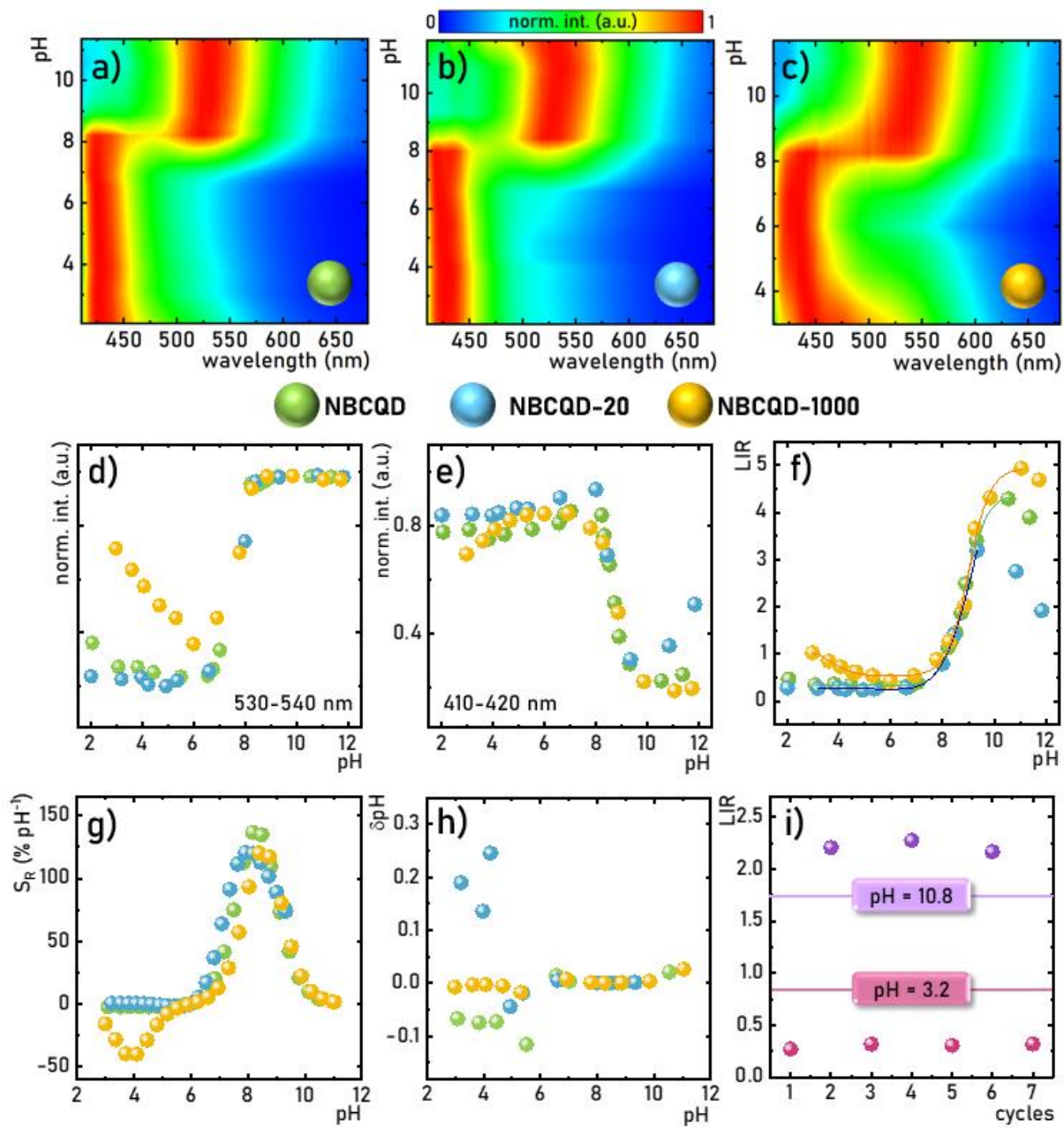

**Figure 4.** Luminescence maps of normalized emission spectra as a function of pH ($\lambda_{exc}$ = 400 nm; T = 20 °C) for colloidal NBCQD (a), NBCQD-20 (b), and NBCQD-1000 (c). Normalized integral emission intensities in the spectral ranges 530-540 nm (d) and 410-420 nm (e), LIR (f), and the corresponding relative sensitivity $S_R$ (g) with pH uncertainty (h) as a function of pH for colloidal NBCQD, NBCQD-20, and NBCQD-1000. LIR repeatability tests (i) - LIR values determined during successive cycles of increasing pH to 10.8 and decreasing pH to 3.2 for colloidal NBCQD-20.

The pH-dependent changes observed in the emission spectra revealed a pronounced color shift, primarily resulting from a drastic change in the intensity ratio between two emission bands located in the blue and green spectral regions, particularly near pH ~8. To visualize this luminescence color evolution, the results were plotted on a CIE 1931 chromaticity diagram (Figure 5a and Figure S5a,c). Analysis of the chromaticity coordinates ($x, y$) as a function of pH (Figure 5c) confirmed that the most significant colorimetric shifts occur within the pH range of approximately 7-10. This trend is especially evident for the NBCQD-20 colloid, where the $x$ coordinate shifts from 0.205 to 0.301 and the $y$ coordinate changes from 0.186 to 0.432 when comparing values at pH = 6.6 and pH = 9.33. These variations closely correlate with the LIR-based calibration curve, reinforcing the consistency between ratiometric LIR-based and colorimetric readouts. This marked chromaticity shift is also clearly visible in the photographs taken under 400 nm excitation (Figure 5b), where a distinct transition in emission color from blue, through turquoise, to green is observed for NBCQD-20 at pH values of 6.6, 7.75, and 8.55. Such visually perceivable changes enabled the development of an alternative, image-based pH sensing approach, based on CIE1931 coordinates, by constructing calibration curves of their values versus pH. Using the same sensitivity equation (Eq. 4), the relative sensitivities were calculated by substituting $x$ or $y$ in place of LIR, and the corresponding Δx or Δy instead of ΔLIR. The resulting pH-dependent $S_{Rx}$ and $S_{Ry}$ values are shown in Figures 5d and 5e. Based on the x-coordinate, the highest sensitivity was achieved for NBCQD, reaching 22.3% $pH^{-1}$ at approximately pH 8.5. For NBCQD-20 and NBCQD-1000, the maximum sensitivities were slightly lower, amounting to 18.8% $pH^{-1}$ at pH ~8.8 and 15.0% $pH^{-1}$ at pH ~8.7, respectively. In contrast, significantly higher sensitivities were obtained when considering the y-coordinate. NBCQD and NBCQD-20 exhibited maximum sensitivities of 46.5% $pH^{-1}$ at pH ~7.7 and 49.5% $pH^{-1}$ at pH ~7.9, respectively. Notably lower maximal sensitivity was recorded for NBCQD-1000, reaching 26.0% $pH^{-1}$ at pH ~7.1. This alternative readout strategy provides a visually intuitive, calibration-compatible, and reproducible method of pH sensing, particularly valuable in applications requiring fast, on-site, or image-based pH estimation without complex instrumentation.

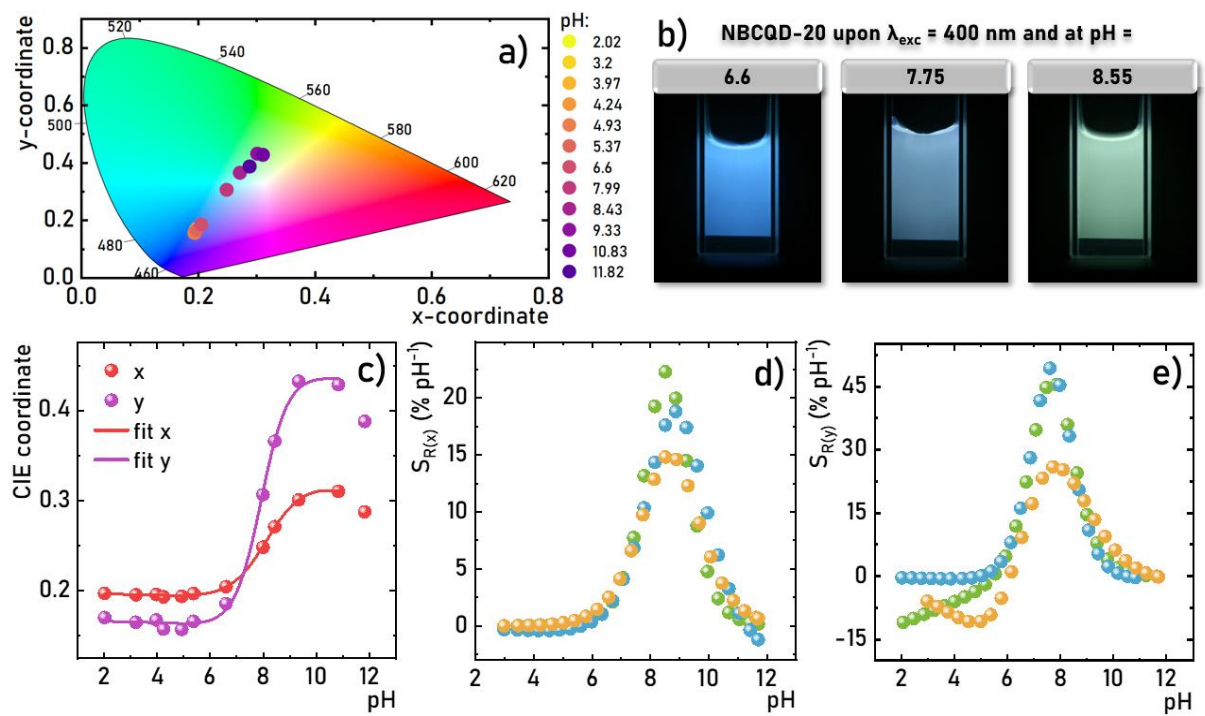

**Figure 5.** CIE 1981 chromaticity diagram for colloidal NBCQD-20 at different pH values upon $\lambda_{exc}$ = 400 nm and T = 20 °C (a). Photographs of colloidal NBCQD-20 at selected pH values (6.6, 7.75, and 8.55) upon $\lambda_{exc}$ = 400 nm (b). CIE y- and x-coordinates as a function of pH for colloidal NBCQD-20 (c), and the corresponding $S_{Rx}$ (d) and $S_{Ry}$ (e) values for NBCQD, NBCQD-20, and NBCQD-1000 colloids.

Current real-life solutions that enable pH monitoring are diverse but also associated with certain limitations. Potentiometric electrodes, such as the widely used glass electrode, can provide accurate and quantitative measurements of pH. However, they are point-probe devices, requiring direct contact with the medium, frequent calibration, and delicate handling, which restricts their use in large-area or real-time mapping. In contrast, chemical indicators and indicator papers offer simple and low-cost alternatives, yet they typically provide only approximate pH values. These methods often rely on visible color changes, which can be subjective and strongly dependent on illumination conditions, observer perception, and sample background. Moreover, indicator papers and dyes are usually single-use, non-reversible, and incapable of delivering precise numerical readouts. Importantly, they can often only confirm whether the pH of a system exceeds or falls below a certain threshold, without offering reliable information on the exact value or allowing continuous monitoring. As a result, there remains a technological gap between high-precision but localized electrode measurements and low-cost but approximate colorimetric methods. This gap highlights the need for pH-sensing strategies that are not only accurate and sensitive but also reversible, reusable, and capable of providing spatially resolved information in real time.

The most commonly used approach for 2D imaging of physical changes based on luminescence is ratiometric imaging, which requires acquiring two images through bandpass filters matched to the spectral ranges defining the calibrated LIR parameter. The pixel intensities of these images are then ratioed, calibrated, and reconstructed into a parameter map. Although this method is attractive due to its straightforward experimental setup - consisting of a camera, an excitation source, and a pair of optical filters - it becomes less effective in environments with rapidly changing conditions. The delay associated with filter switching introduces a temporal mismatch, leading to inaccurate tracking of spectral changes. A more robust strategy is to minimize intermediate acquisition steps, thereby reducing recording time. This can be achieved by directly exploiting the intrinsic RGB sensitivity of the camera.[58] In this way, a single image provides the necessary spectral information without the use of external filters, enabling reliable, real-time monitoring of dynamic changes. However, the effectiveness of this approach relies on spectral changes occurring within the sensitivity range of the RGB channels. As shown in Figure 6a, the emission spectrum of the NBCQD-20 colloid overlaps with the spectral windows of the B and G channels. With increasing pH, the emission shifts toward the G channel, enabling the use of RGB-based imaging. To demonstrate this concept, a proof-of-concept experiment was performed with NBCQD-20 sample. The colloid was adjusted to an acidic pH of 4.2 and deposited as a droplet, which was excited with 400 nm radiation (Figure 6b). A NaOH crystal was then introduced into the droplet, and photographs were taken every 1.5 s, with representative frames presented in Figure 6c and Figure S6. As anticipated, the luminescence evolved from blue to green as the NaOH dissolved. The B and G channels were extracted from the recorded images, and their intensity ratios were used to construct pH distribution maps (Figure 6d and Figure S7). To obtain quantitative information, a calibration curve correlating B/G ratios with pH was established using reference solutions (Figure 6e). This experiment demonstrates that the RGB-based approach enables simultaneous temporal and spatial monitoring of pH. The pH distribution across the droplet (after 21 s), obtained along the cross-section indicated in Figure 6d, reveals the spatial variation of pH (Figure 6f). In parallel, the temporal evolution of the average pH (averaged in the 2x2 mm area in the center of the droplet) at a selected point of the droplet was monitored, providing direct insight into the dynamics of the neutralization process as the NaOH crystal gradually dissolved (Figure 6g). These results highlight the potential of this method, combined with the specifically engineered NBCQD colloids, for reliable, real-time visualization and quantification of pH dynamics. The spectral response of the colloids ensures compatibility with the RGB detection channels, thereby enabling direct, filter-free imaging and establishing them as an effective platform for spatio-temporal pH monitoring.

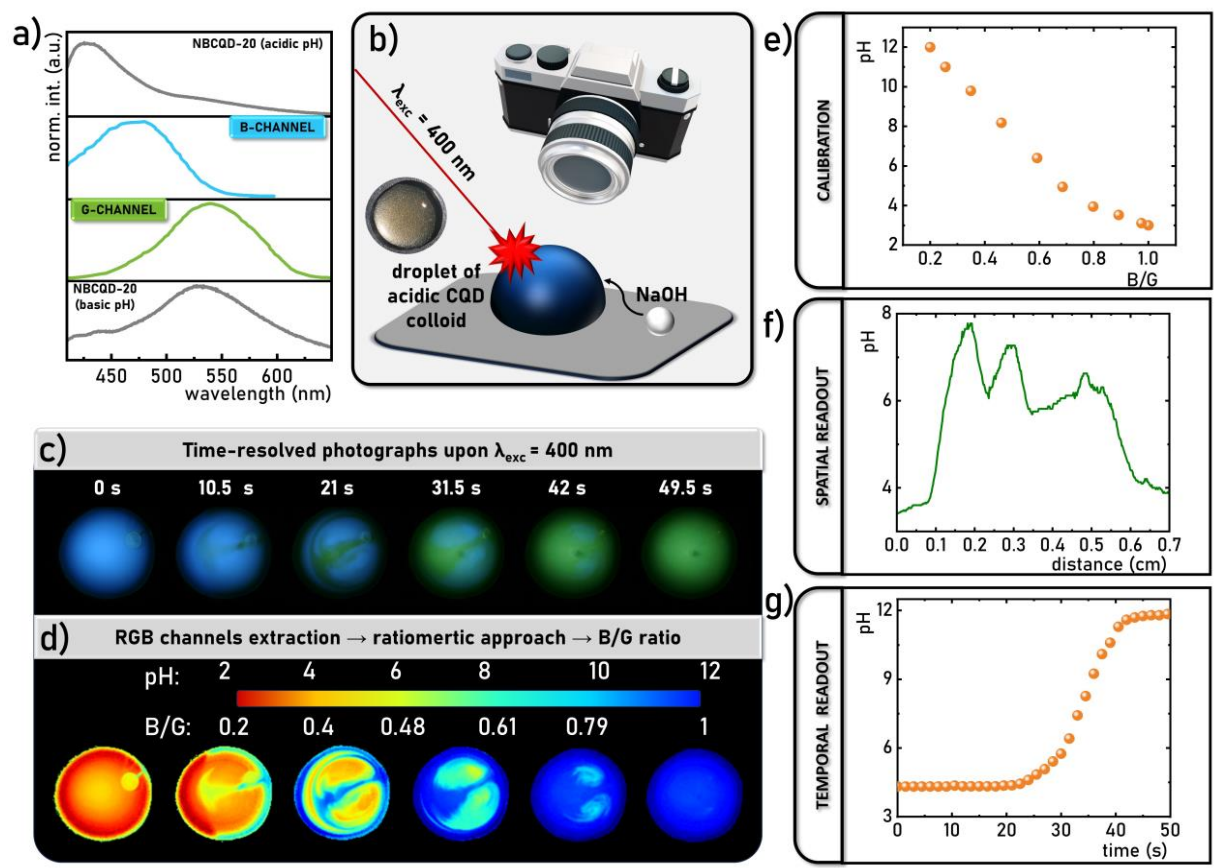

**Figure 6.** Spectral range alignment of the NBCQD-20 colloid ($\lambda_{exc}$ = 400 nm) with the spectral range of the camera's B and G channels (a). Schematic illustration of the proof-of-concept experiment setup (b). Time-lapse photographs of NBCQD-20 colloid droplet showing the evolution of luminescence upon pH increase induced by the addition of a NaOH crystal to the initially acidic colloid (c), together with the corresponding B/G ratio maps (d). Calibration curve of the B/G parameter versus pH (e). pH along the cross-section of the droplet after 21 s (f) and temporal evolution of the average pH at the center of the droplet as a function of time (g).

**Conclusions**

This publication demonstrates the successful application of *m*-aminophenylboronic acid-functionalized carbon quantum dots, synthesized from lignin, for pH monitoring. The luminescence analysis revealed that the obtained NBCQDs exhibit a pronounced and visually perceptible shift in emission color from blue to green with increasing pH. Three complementary sensor design strategies were proposed: based on CIE chromaticity coordinates, the luminescence intensity ratio, and RGB channel extraction. The range of maximum luminescence variation, defining the effective operating window of the sensor, was established between pH ~7 and 10. This range is particularly attractive, as it encompasses near-neutral and slightly basic conditions relevant to environmental, biological, and industrial applications. The developed NBCQD-based sensors exhibited very high sensitivities, reaching 137% $pH^{-1}$ and 49.5% $pH^{-1}$ in the LIR- and CIE-based readouts, respectively, along with exceptionally low measurement uncertainty (<0.01) throughout the operating range. Importantly, the luminescence response was fully reversible, rendering the system reusable and thus economically advantageous compared to disposable probes. A key aspect of this study was the consideration of practical applicability. Through the introduction of an RGB-based readout strategy, pH mapping was enabled in real time using simple instrumentation: UV lamp excitation, standard digital camera, and straightforward image processing. This approach not only dramatically reduces the cost of pH monitoring but also enables spatially and time-resolved measurements - overcoming the limitations of conventional methods such as electrodes, indicator papers, or molecular dyes, which are restricted to localized measurements.

In summary, the results presented in this work highlight the synergy between sustainable nanomaterial design and practical sensing applications. By leveraging renewable biomass precursors and simple, low-cost readout strategies, we demonstrate that eco-friendly lignin-derived carbon quantum dots can serve as highly sensitive and application-ready pH sensors. The proof-of-concept demonstrated herein provides a foundation for the integration of sustainable carbon-based nanomaterials into next-generation sensing technologies with tangible real-world impact.

**Acknowledgements**

This work was supported by National Science Center Poland under project WEAVE-UNISONO No. 2023/05/Y/ST5/00013 and by Slovenian Research Agency (ARIS) through research core funding grants P1-0134, P2-0089 and P1-0418 and research project J2-50061. M. Sz. gratefully acknowledges the support of the Foundation for Polish Science through the START program.